\documentclass[11pt]{article}

\usepackage[margin=2.5cm]{geometry}

\usepackage{newtxtext,newtxmath}
\usepackage{microtype}

\usepackage{xcolor}
\definecolor{arXivBlue}{HTML}{0072B2}
\definecolor{arXivOrange}{HTML}{D55E00}
\definecolor{arXivGreen}{HTML}{009E73}
\definecolor{arXivPurple}{HTML}{CC79A7}

\usepackage[colorlinks=true, allcolors=arXivBlue]{hyperref}

\usepackage{booktabs}
\usepackage{tabularx}
\usepackage{multirow}

\usepackage{graphicx}
\usepackage[font=small,labelfont=bf]{caption}

\usepackage{amsmath}

\usepackage{lineno}
\usepackage{float}

\usepackage{enumitem}
\setlist{nosep}

\title{\textbf{LUMEN: Cost-Transparent Multi-Agent Pipeline for\\Automated Systematic Review and Meta-Analysis}}

\author{
  Yen-Hsun Huang$^1$ \quad Yu-Shiou Lin$^{2,*}$ \\[4pt]
  {\small $^1$Department of Education, Taipei Veterans General Hospital, Taipei, Taiwan} \\
  {\small $^2$Department of Psychiatry, Taipei Veterans General Hospital, Taipei, Taiwan} \\[4pt]
  {\small $^*$Corresponding author: Yu-Shiou Lin, MD. Email: \href{mailto:yslin37@vghtpe.gov.tw}{yslin37@vghtpe.gov.tw}}
}

\date{}

\begin{document}
\maketitle

\begin{abstract}
Systematic reviews and meta-analyses (SR/MA) remain the gold standard for evidence synthesis, yet completing one typically requires 67 weeks and substantial expert effort. Recent large language model (LLM) systems have demonstrated strong performance on individual SR phases---screening (otto-SR: 96.7\% sensitivity), extraction (Gartlehner et al.: 91.0\% accuracy), and search (TrialMind: 0.83 recall)---but no study has reported what it actually costs to run an end-to-end pipeline, how cost distributes across phases, or how architectural choices affect the cost--quality trade-off. We present LUMEN, an open-source multi-agent pipeline that automates six SR/MA phases using 11 specialized LLM agents with deliberate model routing. We evaluate LUMEN on seven datasets: five self-conducted domain reviews (psychiatry, psychology, surgery, vaccinology, cardiology) and two SYNERGY screening benchmarks. Across 13 ground-truth-comparable outcomes, LUMEN achieves 100\% directional agreement with published meta-analyses, with effect sizes within 1\% for homogeneous study designs. The primary contribution is the first empirical cost and operational characterization of such a pipeline: a complete review costs \$19--\$29 (median \$22.65), with title--abstract screening and data extraction together dominating expenditure. A three-arm extraction ablation reveals a phase-dependent architecture reversal: multi-agent design hurts screening but is essential for extraction, producing 5.7$\times$ more poolable analyses than single-model alternatives while eliminating clinically dangerous direction errors. A two-dataset screening benchmark demonstrates that model ranking is domain-dependent and not transferable across review topics. All code and cost logs are publicly available.

\smallskip
\noindent\textbf{Keywords:} systematic review, meta-analysis, large language models, multi-agent systems, cost analysis, automation, open-source
\end{abstract}

\section{Introduction}

Systematic reviews remain the cornerstone of evidence-based medicine, yet their production timeline---a median of 67 weeks across 195 reviews~\cite{borah2017}---creates a widening gap between primary evidence generation and its synthesis. The traditional workflow requires at minimum two independent reviewers for screening, extraction, and quality assessment, with labor costs frequently exceeding tens of thousands of dollars~\cite{michelson2019}. As the volume of primary literature grows, the manual SR/MA model faces a scalability crisis.

Recent LLM-based systems have demonstrated that individual SR phases can be automated with impressive accuracy. Otto-SR achieved 96.7\% screening sensitivity and 93.1\% extraction accuracy across 32,357 citations and 4,559 data points, reproducing 12 Cochrane reviews in two days~\cite{ottosr2025}. TrialMind demonstrated 1.5--2.6-fold screening improvement over baseline methods across 100 SRs and 2,220 clinical studies~\cite{trialmind2025}. Gartlehner et al.\ validated AI-assisted extraction at 91.0\% accuracy across 9,341 data items in a prospective study-within-reviews design~\cite{gartlehner2025}. These systems establish that LLMs can perform SR tasks competently. However, they share three common gaps. First, none reports what it actually costs to run---how many tokens each phase consumes, how model selection affects the cost--quality trade-off, or how cost varies across review domains. Second, most address only a subset of the SR workflow: otto-SR covers screening through meta-analysis but not manuscript generation or quality assessment; TrialMind is validated primarily on oncology; Gartlehner addresses extraction alone. Third, the architectural question of \emph{when} multi-agent design helps versus hurts remains unexamined.

We present LUMEN, an open-source multi-agent pipeline that addresses these gaps. LUMEN automates six phases end-to-end---search strategy, title--abstract screening, full-text review, structured extraction, statistical synthesis, and manuscript drafting---with automated RoB-2/ROBINS-I and GRADE assessment. Each API call is logged with input/output token counts, latency, and model identity, enabling granular cost profiling. The pipeline uses 11 specialized agents with deliberate model routing: high-volume phases use cheaper models (Gemini 3.1 Pro, GPT-4.1 Mini), while high-judgment phases use Claude Sonnet 4.6, and verification uses GPT-5.4.

Our primary contribution is not a claim of state-of-the-art accuracy but the first empirical cost and operational characterization of an end-to-end multi-agent SR/MA pipeline. Specifically, we report: (1)~per-phase cost breakdowns across five clinical domains, (2)~a phase-dependent architecture reversal where multi-agent design hurts screening but is essential for extraction, (3)~cross-dataset evidence that screening model ranking is domain-dependent, and (4)~the observation that cost does not predict quality. All code, cost logs, and prompts are publicly available at \url{https://github.com/YHHuan/LUMEN}.

\section{Methods}

\subsection{Pipeline Architecture}

\begin{figure}
    \centering

    \caption{LUMEN pipeline architecture. \\ Six phases automate the SR/MA workflow using 11 specialized LLM agents with deliberate model routing. Phase~3.1 (expanded) shows the dual-screener mechanism with five-point relevance scoring and arbitration. Median per-phase costs are annotated.}\label{fig:architecture}

    \includegraphics[width=0.9\linewidth]{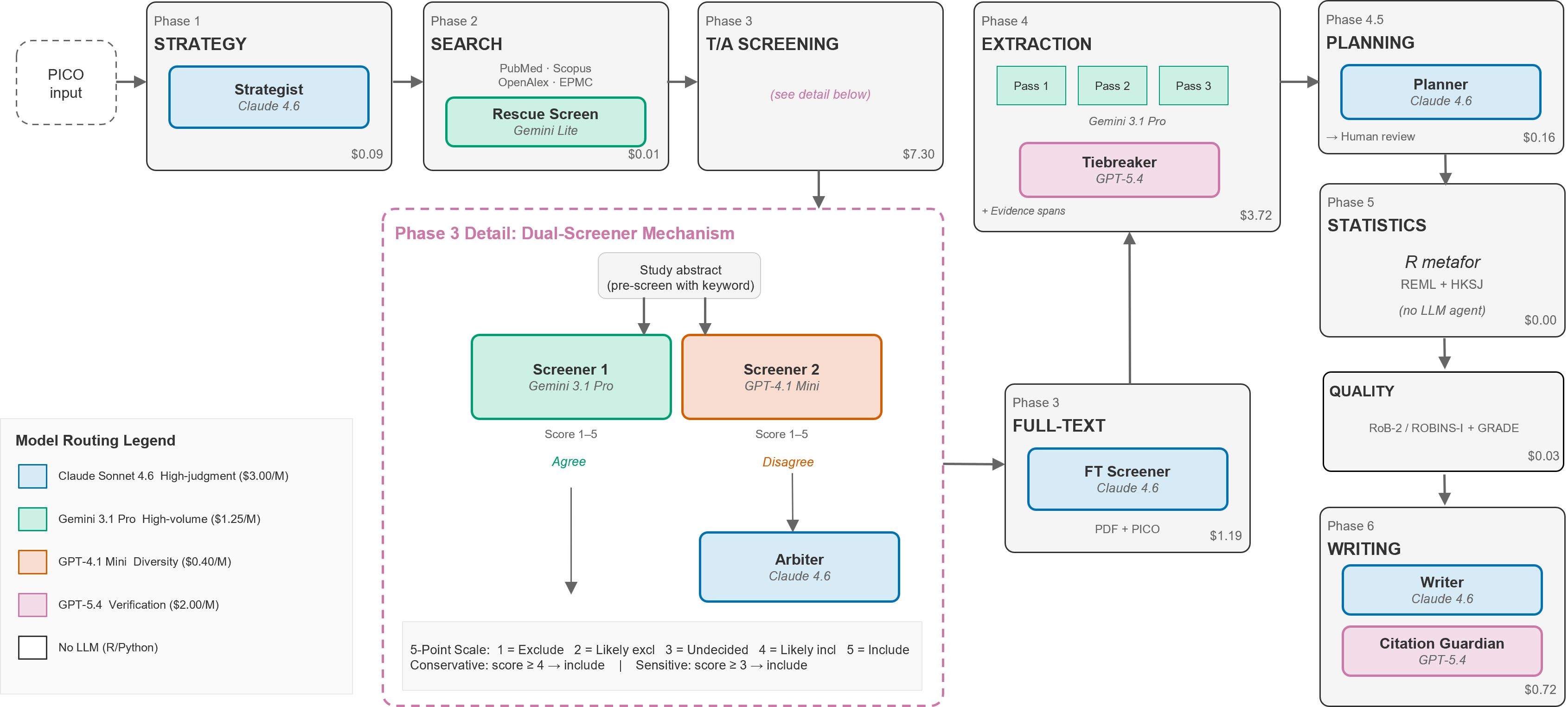}
\end{figure}

LUMEN orchestrates 11 LLM agents across six phases (Figure~\ref{fig:architecture}, Table~\ref{tab:agents}). \textbf{Phase~1} (Strategy) uses Claude Sonnet 4.6 to generate a PICO-derived search strategy and screening criteria. \textbf{Phase~2} searches PubMed, Scopus, OpenAlex, and Europe PMC, followed by deduplication and a pre-screening rescue pass. \textbf{Phase~3.1} (Title--Abstract Screening) employs a dual-screener mechanism: two independent models (Gemini 3.1 Pro and GPT-4.1 Mini) each assign a five-point relevance score using identical static-plus-dynamic prompts, and Claude Sonnet 4.6 arbitrates disagreements with a lean-toward-inclusion policy. \textbf{Phase~3.3} (Full-text Screening) uses Claude Sonnet 4.6 to verify PICO criteria against PDFs. \textbf{Phase~4} (Extraction) applies a three-pass structured extraction using Gemini 3.1 Pro with GPT-5.4 as tiebreaker. \textbf{Phase~4.5} (Planning) profiles extracted data and proposes an analysis plan. \textbf{Phase~5} (Statistics) executes REML random-effects meta-analysis with Knapp--Hartung adjustment via R metafor~\cite{viechtbauer2010}. \textbf{Phase~6} (Writing) generates a citation-grounded manuscript draft.

The prompt architecture follows a static-plus-dynamic design: generic rules are defined in versioned YAML templates, while PICO-specific criteria are injected at runtime. Model routing is deliberate: high-volume phases use cheaper models (\$0.40--\$1.25/M input tokens), high-judgment phases use Claude Sonnet (\$3.00/M), and verification uses GPT-5.4 (\$2.00/M).

\begin{table}[ht]
\centering
\caption{Agent inventory.}\label{tab:agents}
\small
\begin{tabular}{@{}llll@{}}
\toprule
Agent & Phase & Model & Role \\
\midrule
Strategist & 1 & Claude Sonnet 4.6 & PICO $\to$ search strategy \\
Rescue Screener & 2.5 & Gemini Flash Lite & Quarantine rescue \\
Screener 1 & 3.1 & Gemini 3.1 Pro & 5-point T/A screening \\
Screener 2 & 3.1 & GPT-4.1 Mini & 5-point T/A screening \\
Arbiter & 3.1 & Claude Sonnet 4.6 & Conflict resolution \\
Full-text Screener & 3.3 & Claude Sonnet 4.6 & PDF-level PICO verification \\
Extractor & 4 & Gemini 3.1 Pro & 3-pass extraction + evidence spans \\
Tiebreaker & 4 & GPT-5.4 & Cross-pass disagreement resolution \\
Analysis Planner & 4.5 & Claude Sonnet 4.6 & Data profiling $\to$ plan \\
Writer & 6 & Claude Sonnet 4.6 & Citation-grounded drafting \\
Citation Guardian & 6 & GPT-5.4 & Citation marker verification \\
\bottomrule
\end{tabular}
\end{table}

\subsection{Cost and Operational Tracking}

Every API call is logged locally in structured JSON: model identity, input/output token counts, call latency, retry count, and the initiating phase and agent. Cost is computed post hoc by multiplying token counts by per-model pricing at execution time. All cost logs are included in the public repository.

\subsection{Validation Design}

Validation establishes that outputs meet a sufficient quality threshold, ensuring the cost analysis is conducted on a pipeline producing meaningful results.

\textbf{Screening validation} uses two SYNERGY benchmark datasets with expert-labeled ground truth, eliminating developer-as-evaluator bias. Four LLM arms---single Gemini, single GPT, single Claude, and full LUMEN dual+arbiter---are compared at two operating points: \emph{conservative} (score~$\geq$~4) and \emph{sensitive} (score~$\geq$~3). The five-point scale produces a natural ``undecided'' zone (score~=~3) whose handling determines the sensitivity--specificity trade-off; reporting both thresholds avoids selective reporting and lets practitioners choose the operating point appropriate to their tolerance for false negatives.

\textbf{Extraction validation} uses a three-arm ablation across five domains: Arm~A (LUMEN 3-pass), Arm~C (single-pass Sonnet), Arm~D (single-pass Gemini), evaluated on poolability, directional agreement, and clinically dangerous errors. Arm~B (GPT single-pass) is omitted because GPT-5.4 serves as the tiebreaker in Arm~A, making it non-independent; the ablation tests architecturally independent alternatives only.

\textbf{Synthesis validation} compares LUMEN-derived pooled estimates against published ground truth on direction and magnitude.

\subsection{Datasets}

Seven datasets are used (Table~\ref{tab:datasets}). Two SYNERGY benchmarks provide external screening validation. Five self-conducted domain reviews provide material for cost profiling, extraction ablation, and synthesis comparison.

\begin{table}[ht]
\centering
\caption{Dataset characteristics.}\label{tab:datasets}
\small
\begin{tabular}{@{}llrrrl@{}}
\toprule
ID & Domain & Yield & Screened & Included & GT Reference \\
\midrule
D1 & Antidepressants in elderly & 5,168 & 3,756 & 16 & Lenouvel 2024 \\
D2 & CBT-I for insomnia + depression & 5,535 & 3,172 & 10 & Furukawa 2024 \\
D3 & Lap.\ vs open cholecystectomy & 6,233 & 4,334 & 46 & Roy 2024 \\
D4 & Pneumococcal vaccines & 2,010 & 1,389 & 49 & Farrar 2023 \\
D5 & SGLT2i in heart failure & 8,062 & 4,839 & 11 & Vaduganathan 2022 \\
S1 & SYNERGY: Bos\_2018 & --- & 4,757 & 10 & Expert labels \\
S2 & SYNERGY: van\_de\_Schoot\_2018 & --- & 4,314 & 38 & Expert labels \\
\bottomrule
\end{tabular}
\end{table}

\section{Results}

\subsection{Quality Validation}

LUMEN completed all pipeline phases across five domain reviews. Across 13 ground-truth-comparable outcomes, the pipeline achieved \textbf{100\% directional agreement} with published meta-analyses. Effect magnitude concordance was strongest in homogeneous study designs: Domain~5 (SGLT2 inhibitor mega-trials, 11 included studies) reproduced all four published hazard ratios within 1\% (primary composite HR~0.777 vs 0.77; heart failure hospitalization HR~0.724 vs~0.72), while Domain~3 (cholecystectomy RCTs) matched all four outcome directions with $I^2$ near 0\%. In heterogeneous domains (D1, D2, D4), effect sizes agreed directionally but $I^2$ was higher, reflecting outcome standardization gaps requiring human clinical judgment (Section~4.2). A forest plot comparison of select outcomes is in the Supplement (Figure~S1).

Inter-rater agreement between dual screeners ranged from $\kappa$~=~0.62 to 0.76 (PABAK 0.93--0.99). Having established sufficient quality, we turn to the primary contribution.

\subsection{Cost Anatomy}

Total cost ranged from \$19.51 (D3, 46 included studies) to \$29.04 (D5, 11 included studies), with a \textbf{median of \$22.65} (Table~\ref{tab:cost}). Screening and extraction together dominated expenditure, but their relative share varied markedly: in high-yield, low-inclusion domains (D1, D2, D5), title--abstract screening consumed 74--86\% of total cost, while in low-yield, high-inclusion domains (D3, D4), extraction consumed 47--49\%. All remaining phases combined contributed 4--22\% depending on domain.

\begin{table}[ht]
\centering
\caption{Cost breakdown by phase (USD).}\label{tab:cost}
\small
\begin{tabular}{@{}lrrrrrr@{}}
\toprule
Phase & D1 & D2 & D3 & D4 & D5 & Range \\
\midrule
P3.1 T/A Screening & 20.37 & 16.80 & 5.69 & 7.30 & 25.00 & \$5.69--\$25.00 \\
P4 Extraction & 3.72 & 3.69 & 9.50 & 10.67 & 2.92 & \$2.92--\$10.67 \\
P3.3 Full-text & 0.66 & 1.19 & 2.41 & 2.70 & 0.04 & \$0.04--\$2.70 \\
P6 Manuscript & 0.79 & 0.41 & 0.72 & 0.74 & 0.62 & \$0.41--\$0.79 \\
Other$^\dagger$ & 0.56 & 0.42 & 1.19 & 1.24 & 1.08 & \$0.42--\$1.24 \\
\midrule
\textbf{Total} & \textbf{26.10} & \textbf{22.51} & \textbf{19.51} & \textbf{22.65} & \textbf{29.04} & \textbf{\$19.51--\$29.04} \\
\bottomrule
\end{tabular}

{\footnotesize $^\dagger$Other = Strategy + Pre-screen + Planning + Quality (RoB/GRADE).}
\end{table}

\begin{figure}[htbp]
    \centering
    \caption[Cost breakdown]{Cost breakdown \\ 
    \small Per-phase cost breakdown across five domain reviews. Title--abstract screening (blue) dominates in high-yield domains (D1, D2, D5), while extraction (orange) dominates in high-inclusion domains (D3, D4). Dashed line indicates median total cost (\$22.65).}
    
    \includegraphics[width=0.9\linewidth]{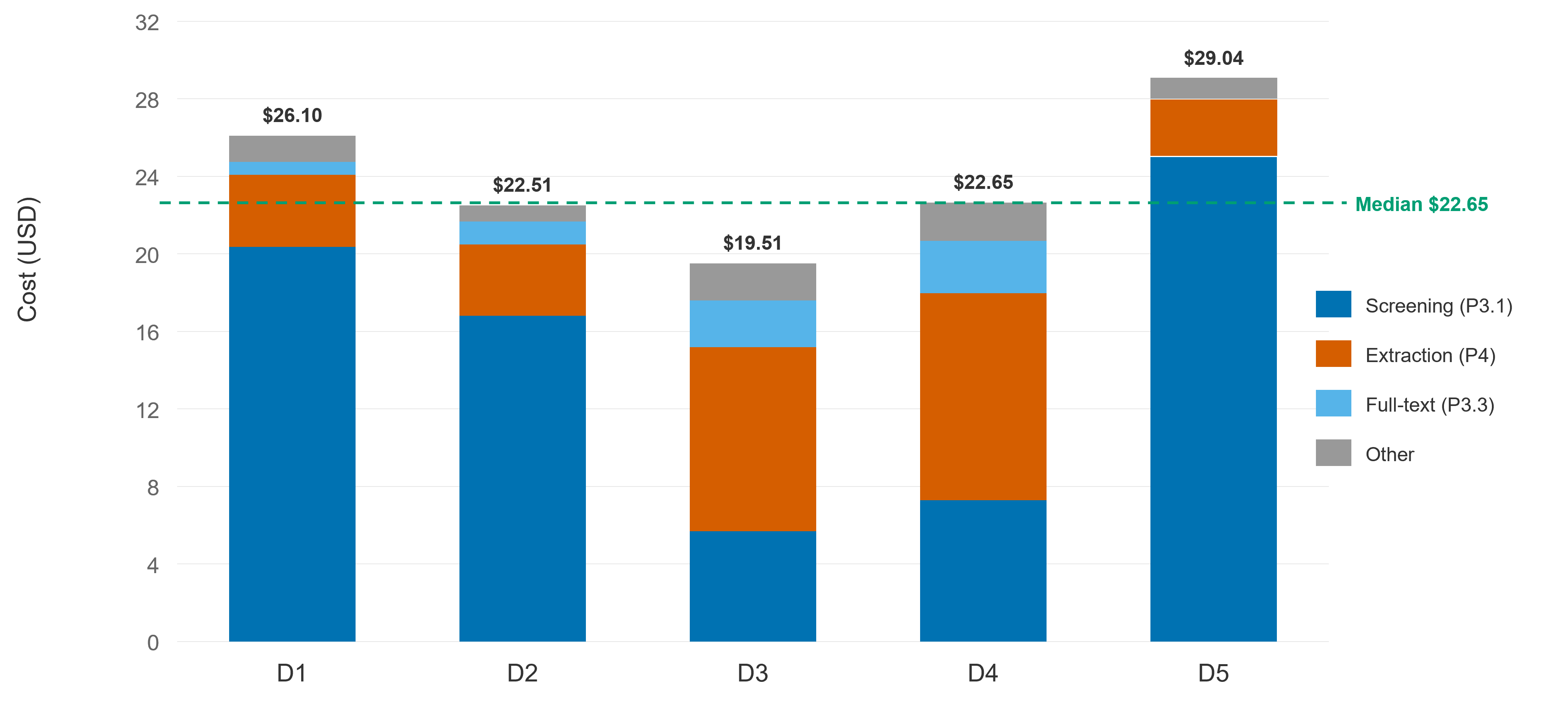}
    \label{fig:cost}
\end{figure}

Screening cost scales linearly with deduplicated study count (D5: 4,839 studies at \$25.00), while extraction scales with included studies and full-text length (D4: 49 studies at \$10.67). Which phase dominates depends on the review's \emph{yield-to-inclusion ratio}: high ratios (D1: 3,756/16~=~235:1) make screening dominant; low ratios (D3: 4,334/46~=~94:1) shift dominance to extraction. Wall-clock time ranged from 5.1~hours (D4) to 10.8~hours (D1), with zero retries and zero failures across all runs (Table~S1).

\subsection{Phase-Dependent Architecture}

The central architectural finding is a phase-dependent reversal: multi-agent design hurts screening but is essential for extraction.

\paragraph{Screening: single~$\geq$~multi.}
The SYNERGY benchmark across two datasets confirmed that single-model arms match or exceed dual-screener sensitivity (Table~\ref{tab:screening}). At the sensitive threshold (score~$\geq$~3, absorbing undecided cases), Claude and Gemini both achieved~0.90 on Bos\_2018 and~0.947 on van\_de\_Schoot\_2018, while GPT matched this on Bos (0.90) but dropped to the worst performer on van\_de\_Schoot (0.842, missing 6 of 38 studies). The dual mechanism matched the best single-model sensitivity but did not exceed it; its value lies in producing a structured disagreement queue (22--28\% of studies flagged) that provides an auditable human review layer. Figure4 plots sensitivity against review burden across both datasets: Claude is the only arm that moves toward the ideal trade-off (high sensitivity, low review) on both, achieving~0.947 with only 5\% flagged on van\_de\_Schoot versus 17--24\% for other arms.

\begin{table}[ht]
\centering
\caption{Screening benchmark at the sensitive operating point (score~$\geq$~3).}\label{tab:screening}
\small
\begin{tabular}{@{}llccccr@{}}
\toprule
Dataset & Arm & Sensitivity & Specificity & Missed & Review~\% & Cost \\
\midrule
\multirow{4}{*}{Bos\_2018 (10~GT)} & Gemini & 0.800~(8/10) & 0.649 & 2 & 24 & \$2.36 \\
& GPT & \textbf{0.900}~(9/10) & 0.700 & 1 & 19 & \$0.33 \\
& Claude & \textbf{0.900}~(9/10) & 0.691 & 1 & 17 & \$34.71 \\
& Dual & \textbf{0.900}~(9/10) & 0.618 & 1 & 28 & \$2.88 \\
\midrule
\multirow{4}{*}{\shortstack[l]{van\_de\_Schoot\\(38~GT)}} & Gemini & \textbf{0.947}~(36/38) & 0.773 & 2 & 22 & \$17.53 \\
& GPT & 0.842~(32/38) & 0.824 & 6 & 17 & \$2.40 \\
& Claude & \textbf{0.947}~(36/38) & \textbf{0.940} & 2 & \textbf{5} & \$25.51 \\
& Dual & \textbf{0.947}~(36/38) & 0.758 & 2 & 24 & \$20.02 \\
\bottomrule
\end{tabular}
\end{table}

\begin{figure}
    \centering
    \caption{Screening scatter\\Screening sensitivity versus human review burden at the sensitive operating point (score~$\geq$~3) across four LLM arms on two SYNERGY datasets. Filled markers: Bos\_2018; open markers: van\_de\_Schoot\_2018; arrows trace cross-dataset movement. Upper-right = ideal (high sensitivity, low review). Claude (orange) is the only arm moving toward this ideal on both datasets. GPT (green) reverses from 0.90 to 0.84, missing 6 of 38 studies. Dual (purple) maintains top-tier sensitivity; its higher review\% reflects a structured disagreement queue.}
    \includegraphics[width=0.9\linewidth]{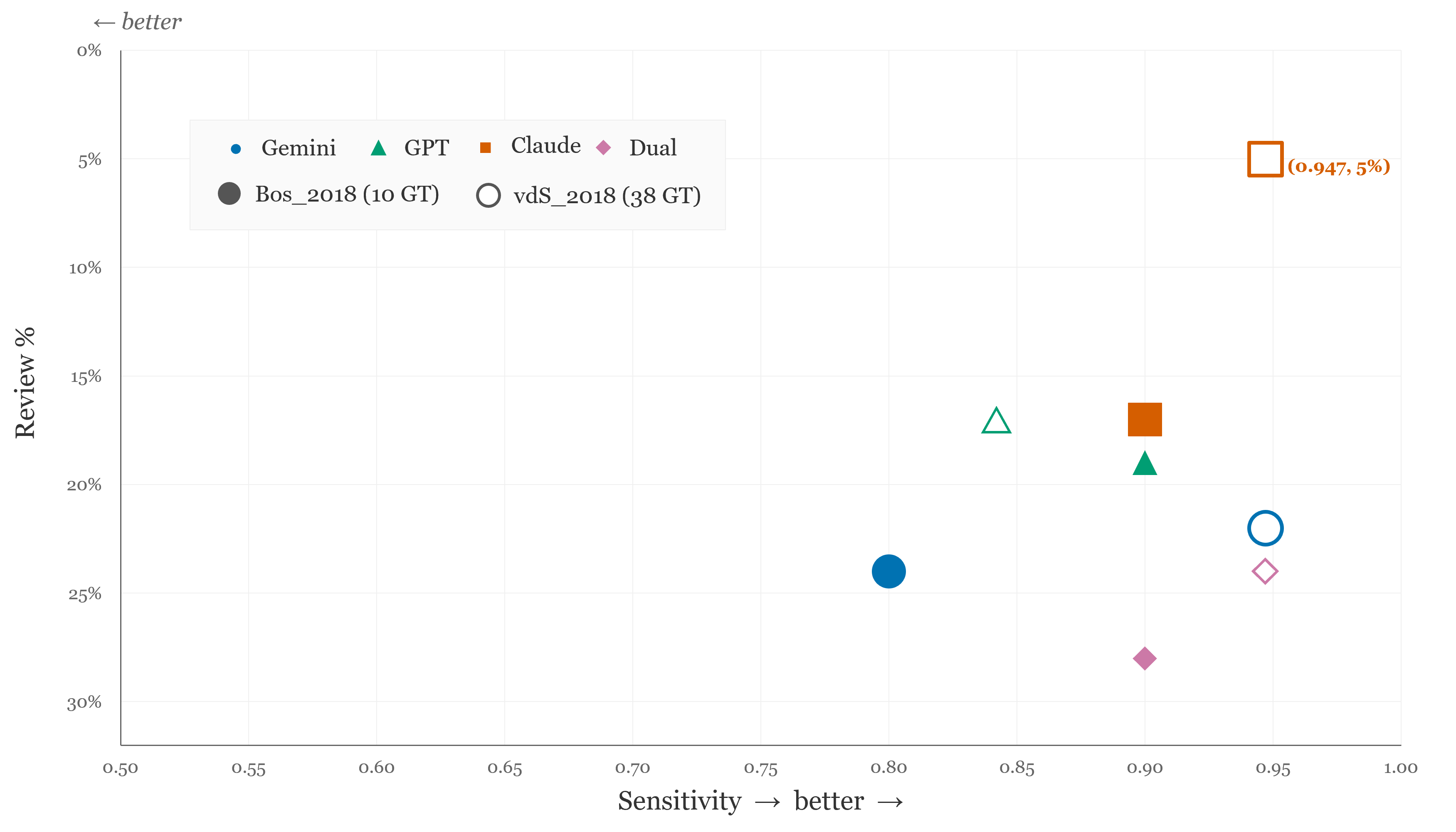}
    \label{fig:screening_scatter}
\end{figure}

This cross-dataset reversal in GPT ranking demonstrates that \textbf{screening model performance is domain-dependent and not transferable}. Claude was the only model with consistent top-tier performance across both datasets, though at 10--100$\times$ the cost. One study per dataset was missed by all four independent arms at both thresholds---a convergent false negative suggesting that these studies sit at the genuine boundary of inclusion criteria rather than representing systematic LLM failure. At the stricter conservative threshold (score~$\geq$~4), sensitivity dropped substantially on van\_de\_Schoot (all arms below~0.76); full two-threshold results are in Table~S5.

\paragraph{Extraction: multi~$\gg$~single.}
The three-arm ablation tells the opposite story (Table~\ref{tab:ablation}). Arm~A produced \textbf{34 poolable analyses} versus 22 (Arm~D) and 6 (Arm~C), with 100\% directional agreement (13/13). Arm~D had one clinically dangerous direction error (D3 biliary: OR~=~0.88 vs correct 2.15). Arm~C failed in two of five domains due to verbatim outcome naming defeating harmonization.

\begin{table}[ht]
\centering
\caption{Extraction ablation: poolable analyses and GT agreement.}\label{tab:ablation}
\small
\begin{tabular}{@{}lccccc@{}}
\toprule
Domain & Arm A & Arm C & Arm D & GT (A) & Cost ratio \\
\midrule
D1 Antidepressants & 6 & 0 & 4 & 1/1 $\checkmark$ & 2.8$\times$ \\
D2 CBT-I & 4 & 0 & 0 & 1/1 $\checkmark$ & 3.9$\times$ \\
D3 Cholecystectomy & 11 & 2 & 8$^*$ & 4/4 $\checkmark$ & 2.8$\times$ \\
D4 Vaccines & 6 & 1 & 4 & 3/3 $\checkmark$ & 2.6$\times$ \\
D5 SGLT2i & 7 & 3 & 6 & 4/4 $\checkmark$ & 3.0$\times$ \\
\midrule
\textbf{Total} & \textbf{34} & \textbf{6} & \textbf{22} & \textbf{13/13} & \textbf{2.6--3.9$\times$} \\
\bottomrule
\end{tabular}

{\footnotesize $^*$D3 Arm D: one clinically dangerous direction error (biliary OR~=~0.88 vs correct 2.15).}
\end{table}

\begin{figure}
    \centering
    \caption{Extraction ablation\\Number of poolable meta-analyses produced by each extraction arm. The LUMEN 3-pass pipeline (Arm~A, blue) consistently produces the most analyses with 100\% GT directional agreement. Arm~C (Sonnet, orange) fails in D1 and D2 due to verbatim naming. Asterisk on D3 Arm~D indicates a clinically dangerous direction error.}
    \includegraphics[width=1.0\linewidth]{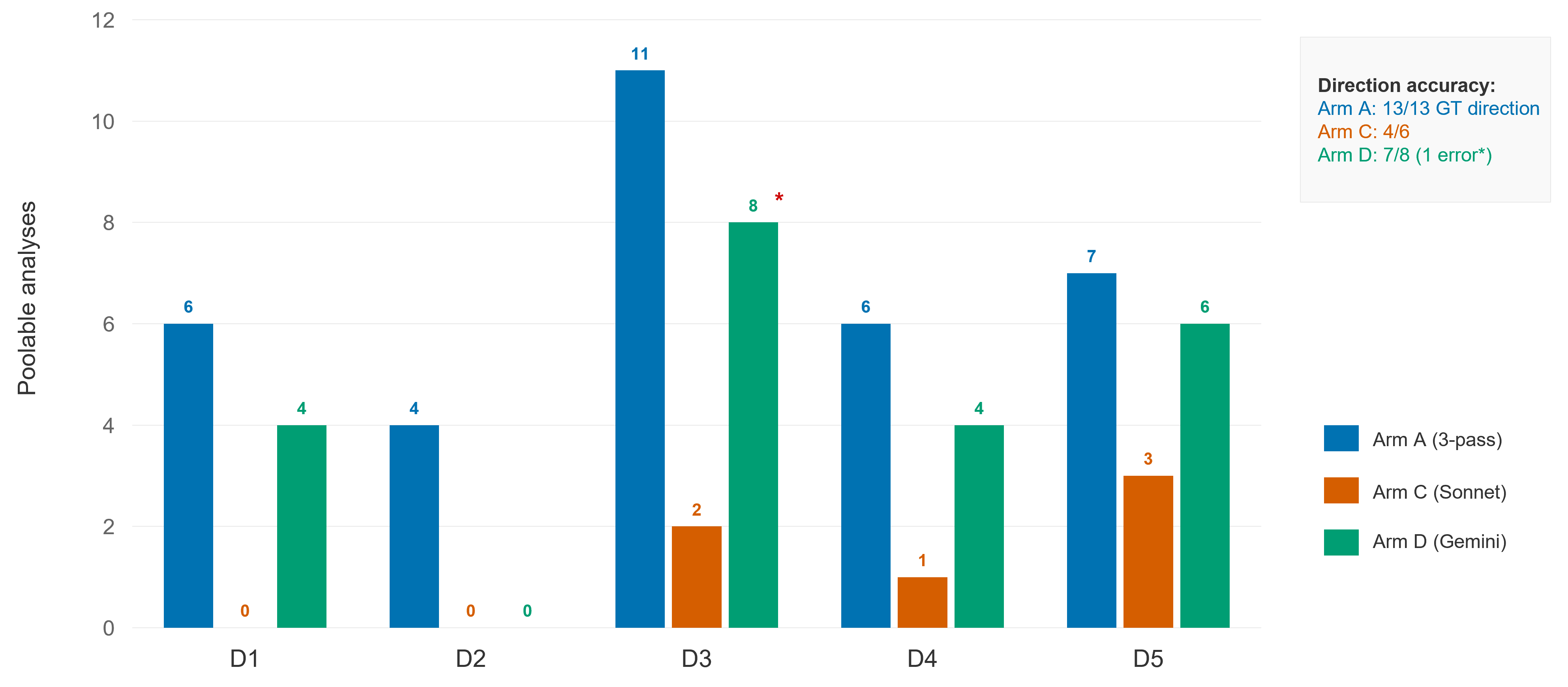}
    \label{fig:ablation}
\end{figure}

The cost premium of 2.6--3.9$\times$ is justified by the 5.7$\times$ poolability gain over Sonnet and the elimination of direction errors that propagate silently into pooled estimates. The dual mechanism's value in screening is not sensitivity but PRISMA-compliant auditability: it generates a disagreement-ranked human review queue that documents where models diverged and why the arbiter intervened. The reversal implies that multi-model budgets should be invested in extraction, not screening.

\subsection{Cross-Domain Cost Variability}

Cost varied 1.5-fold (\$19.51--\$29.04), driven by search yield and included study count. Notably, cost did not predict quality: D3, the cheapest (\$19.51) despite having the most included studies (46), achieved 100\% directional agreement on all four outcomes. D5, the most expensive (\$29.04) due to its large screening corpus, achieved the best effect-size concordance (all HRs within 1\%). D1, the second most expensive (\$26.10), had the highest $I^2$ divergence. The primary quality determinant was study design homogeneity, not cost invested. At the per-finding level, \$22.65 across 13 outcomes yields $\sim$\$1.74 per directionally-correct clinical finding.

\section{Discussion}

\subsection{Cost Positioning}
 
LUMEN completes a full SR/MA in 3.5--10.8 hours at a median of \$22.65, versus a traditional median of 67 weeks~\cite{borah2017}. The cost structure is transparent and decomposable, enabling prospective estimation. Otto-SR~\cite{ottosr2025} achieves 93.1\% extraction accuracy and reproduced 12 Cochrane reviews in two days but reports no cost data and omits manuscript generation and quality assessment. TrialMind~\cite{trialmind2025} validated across 100 SRs but is limited to oncology. Gartlehner et al.~\cite{gartlehner2025} demonstrated 91.0\% accuracy with 0.8\% fabrication but addresses extraction alone. LUMEN contributes an orthogonal dimension: the first empirical answer to what end-to-end automation costs.
 
The comparison with human labor requires careful framing. LUMEN's \$19--\$29 captures only API expenditure; it does not account for researcher time spent defining the PICO, validating outputs, and interpreting results. A fairer characterization is that LUMEN compresses the mechanical phases into hours while leaving the intellectual phases---the actual scholarly contribution---to the researcher. The relevant comparison is not ``\$20 versus \$50{,}000'' but ``months of screening and extraction labor versus hours of validation and interpretation.''
 
The absolute dollar figures are time-stamped to March 2026 pricing. However, the structural finding---that cost distributes between screening and extraction in proportions governed by the yield-to-inclusion ratio---reflects pipeline architecture, not pricing, and should remain stable across pricing regimes.

\subsection{Architectural Implications}
 
The phase-dependent reversal is a design principle: adding agents uniformly is suboptimal. For screening, a single high-capability model outperforms the multi-agent committee (Claude: 0.90--0.95 sensitivity, 5\% review burden). The task structure explains why: title-abstract screening is a rapid relevance judgment where introducing a second screener and arbitrator amplifies disagreement in the ambiguous middle zone rather than adding complementary information.
 
Extraction presents the opposite profile. Study data is embedded in heterogeneous formats---tables, narrative text, supplementary files. A single pass frequently produces direction errors, naming inconsistencies, and unit mismatches. The 3-pass mechanism catches these at the structural level, justifying the 2.6--3.9$\times$ cost premium. LUMEN's evidence-span mechanism provides additional safeguard against fabrication, contrasting with the 0.8\% rate in~\cite{gartlehner2025}.
 
An unexpected finding is that model naming style is a first-order determinant of downstream poolability---the same intervention appearing as three distinct entries fragments what should be a single forest plot. Extraction prompts should enforce canonical nomenclature, a lesson applicable to any LLM-based data extraction pipeline.

\subsection{The Heterogeneity Problem}
 
$I^2$ diverged most from ground truth in heterogeneous outcome domains (D1, D2, D4) and least in homogeneous designs (D3, D5). These are not pipeline defects---a human performing the same mechanical steps would observe identical heterogeneity. The divergence arises because human meta-analysts reshape the analysis using domain knowledge. In Domain~2, the reference review recoded continuous sleep-quality scores into binary response categories ($>$50\% improvement) specifically to reduce heterogeneity---a clinical judgment that no extraction algorithm can replicate. In Domain~5, RCT-derived event rates leave little room for heterogeneity to accumulate.
 
This boundary applies to any automated synthesis system: heterogeneity management is not a feature to be engineered but a constraint of the meta-analytic method itself. LUMEN faithfully executes the mechanical steps; the decision to recode, subgroup, or abandon pooling remains a human responsibility.

\subsection{Screening Subjectivity and the Inclusion Boundary}
 
Disagreement between LUMEN and ground truth often reflects genuine ambiguity rather than error. Screening criteria are written in natural language, and reasonable reviewers routinely disagree on borderline cases---a trial may spawn 30 sub-analyses, only some within a given review's scope. We therefore report two operating points: conservative (undecided excluded, maximizing specificity) and sensitive (undecided included, maximizing recall), letting researchers select the threshold matching their tolerance for false inclusions versus missed studies.
 
The broader question is what screening should look like when computation is cheap. The traditional assumption that screening must be exhaustively sensitive made sense when including a marginal study cost weeks of additional work. When full-text review costs cents, a workflow that accepts slightly lower specificity in exchange for dramatically reduced human burden may be more rational. LUMEN's 5\% human-review queue suggests this trade-off is already practical.

\subsection{Cross-Domain Comparability}
 
The five domains differ in literature volume, publication era, outcome heterogeneity, and ground-truth construction. Placing them side by side and comparing sensitivity or $\kappa$ implicitly assumes comparable difficulty, which is false. We therefore avoid cross-domain averages throughout this paper: all metrics are reported per domain, and where summary statistics are unavoidable, we report ranges rather than means. A sensitivity of 0.85 in a domain with ambiguous criteria is not meaningfully worse than 0.95 in a domain with clear-cut RCTs. The relevant question is whether the pipeline performs adequately \textit{within each domain given its characteristics}.

\subsection{Limitations}
 
Six limitations: (1)~Pre-digital literature bottleneck---Domain~3 lost 9/15 GT studies to PDF unavailability before 2000. (2)~GT comparisons inherit reference review bias; partially mitigated by SYNERGY. (3)~Results specific to March 2026 model versions/pricing. (4)~No automated PICO refinement; the complexity of real clinical literature---ambiguous trial names, overlapping interventions, inconsistent outcome definitions---means PICO specification resists full automation. Assistive refinement is a natural next step. (5)~Non-clinical generalizability untested. (6)~Extraction validation uses poolability concordance rather than per-field accuracy, precluding direct comparison with otto-SR's 93.1\%.

\subsection{The \$20 Question}
 
What does \$20--\$30 buy? Correct directional findings on all comparable outcomes, 34 poolable analyses with evidence spans, automated RoB and GRADE assessments, and a citation-grounded manuscript draft. What it does not buy: outcome recoding, clinical subgroup decisions, or heterogeneity interpretation.
 
This points to a broader reflection. Systematic review is often presented as objective evidence synthesis---a mechanical process yielding reproducible conclusions. Our experience suggests otherwise: from PICO definition to screening thresholds to outcome harmonization, the process is shaped by researcher judgment at every stage. Meta-analysis is not the objective endpoint it is sometimes portrayed as; it is a structured framework within which expert judgment operates. The value of a \$20 pipeline is not removing human judgment but making the mechanical steps cheap enough to redirect researcher time toward the decisions that actually matter. The pipeline is a structured, auditable starting point---not a finished product, but a foundation on which domain expertise builds the final analysis.

\section{Conclusion}
 
We presented LUMEN, the first open-source, cost-transparent multi-agent pipeline for end-to-end SR/MA automation. Three findings stand out: a complete review costs \$19--\$29 with screening and extraction dominating in predictable, domain-dependent proportions; multi-agent architecture exhibits a phase-dependent reversal (harmful for screening, essential for extraction); and screening model performance is domain-dependent with rankings reversing across datasets. What \$20 buys is not a finished meta-analysis but a structured, auditable starting point that lets researchers decide where human expertise is most needed. All code and data are available at \url{https://github.com/YHHuan/LUMEN}.


\clearpage

\appendix
\renewcommand{\thetable}{S\arabic{table}}
\setcounter{table}{0}
\renewcommand{\thefigure}{S\arabic{figure}}
\setcounter{figure}{0}

\section*{Supplementary Material}

\begin{figure}[H]
    \centering
    \caption{Forest plot\\Forest plot comparison of LUMEN-derived pooled estimates (blue) versus published ground truth (orange) for select outcomes in Domains~3 and~5. All estimates agree directionally, with Domain~5 hazard ratios within 1\% of published values.}
    \includegraphics[width=1.0\linewidth]{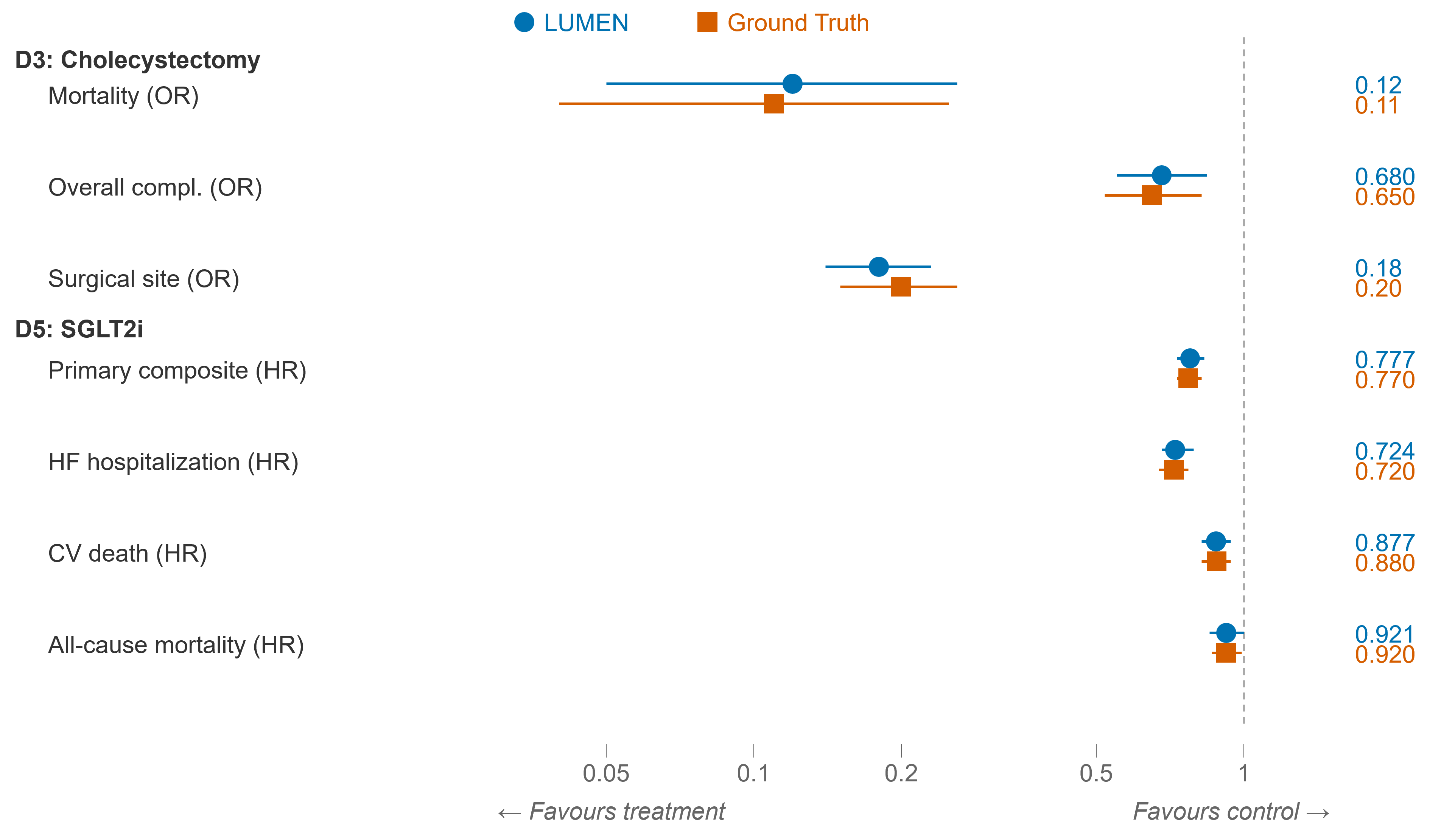}
    \label{fig:forest}
\end{figure}

\newpage

\begin{table}[ht]
\centering
\caption{Per-phase operational metrics aggregated across five domain reviews. Median values reported with range in parentheses. Zero retries and zero failures across all five runs ($\sim$27{,}430 total API calls).}\label{tab:s1}
\small
\begin{tabular}{@{}lrrrrc@{}}
\toprule
Phase & API Calls & Input (M tok.) & Output (K tok.) & Wall-Clock (min) & Fail \\
\midrule
P1 Strategy        & 2           & 0.008          & 5.0        & 0.4          & 0\% \\
P2.5 Pre-screen    & 56 (14--80) & 0.095          & 2.5        & 2.0          & 0\% \\
P3.1 T/A Screen    & 2{,}352 (1{,}748--7{,}520) & 5.16 & 288  & 242          & 0\% \\
P3.3 FT Screen     & 41 (1--86)  & 0.457          & 9.7        & 18           & 0\% \\
P4 Extraction      & 48 (3--147) & 0.434          & 338        & 54           & 0\% \\
P4.5 Planning      & 2 (2--3)    & 0.011          & 7.8        & $<$2         & 0\% \\
P6 Manuscript      & 289 (138--341) & 0.114       & 33.6       & 15           & 0\% \\
Quality            & 17 (4--48)  & 0.012          & 28.6       & 5.9          & 0\% \\
\bottomrule
\end{tabular}
\end{table}

\begin{table}[ht]
\centering
\caption{Phase~3.1 dual-screener agreement metrics. PABAK = prevalence-adjusted bias-adjusted kappa. D5 $\kappa = 0.21$ reflects a kappa paradox ($\sim$97\% exclude); PABAK $= 0.93$ corrects for prevalence.}\label{tab:s2a}
\small
\begin{tabular}{@{}lrrrrrr@{}}
\toprule
Domain & Screened & Included & Human Rev. & Cohen's $\kappa$ & PABAK & Arbiter \\
\midrule
D1 & 3{,}756 & 22  & 49  & 0.74 & 0.99 & 8  \\
D2 & 3{,}172 & 20  & 55  & 0.76 & 0.99 & 5  \\
D3 & 4{,}334 & 185 & 296 & 0.73 & 0.95 & 70 \\
D4 & 1{,}389 & 45  & 51  & 0.62 & 0.94 & 26 \\
D5 & 4{,}839 & --- & 309 & 0.21 & 0.93 & 54 \\
\bottomrule
\end{tabular}
\end{table}

\begin{table}[ht]
\centering
\caption{Five-point confidence distributions for Screener~1 (Gemini 3.1 Pro, top) and Screener~2 (GPT-4.1 Mini, bottom). GPT uses score~2 far more often (D3: 906 vs.\ 114), indicating a more conservative exclusion style.}\label{tab:s2bc}
\small

\medskip
\textbf{Screener 1 --- Gemini 3.1 Pro}\\[4pt]
\begin{tabular}{@{}lrrrrr@{}}
\toprule
Score & D1 & D2 & D3 & D4 & D5 \\
\midrule
1 -- Most likely exclude & 3{,}680 & 3{,}033 & 3{,}745 & 1{,}282 & 4{,}454 \\
2 -- Likely exclude       & 14      & 71      & 114     & 18      & 61      \\
3 -- Undecided            & 42      & 49      & 273     & 48      & 287     \\
4 -- Likely include       & 2       & 3       & 26      & 1       & 4       \\
5 -- Most likely include  & 18      & 16      & 176     & 40      & 33      \\
\bottomrule
\end{tabular}

\bigskip
\textbf{Screener 2 --- GPT-4.1 Mini}\\[4pt]
\begin{tabular}{@{}lrrrrr@{}}
\toprule
Score & D1 & D2 & D3 & D4 & D5 \\
\midrule
1 -- Most likely exclude & 3{,}530 & 2{,}872 & 3{,}039 & 1{,}217 & 4{,}260 \\
2 -- Likely exclude       & 172     & 245     & 906     & 84      & 310     \\
3 -- Undecided            & 23      & 27      & 133     & 10      & 100     \\
4 -- Likely include       & 3       & 5       & 67      & 26      & 16      \\
5 -- Most likely include  & 28      & 23      & 189     & 52      & 153     \\
\bottomrule
\end{tabular}
\end{table}

\begin{table}[ht]
\centering
\caption{Extraction ablation per-analysis results. Arm~A = LUMEN 3-pass; Arm~C = single-pass Sonnet; Arm~D = single-pass Gemini. ``---'' = no poolable analysis. $^*$Direction error (OR=0.88 vs.\ correct 2.15).}\label{tab:s3}
\small
\begin{tabular}{@{}llll@{}}
\toprule
Analysis & Arm A & Arm C & Arm D \\
\midrule
\multicolumn{4}{@{}l}{\textbf{D1: Antidepressants in Elderly}} \\
Depression (all AD)     & $k$=8, SMD=0.12, $I^2$=59\%    & --- & --- \\
Depression (SSRIs)      & $k$=6, SMD=$-$0.02, $I^2$=51\% & --- & --- \\
Response/remission      & $k$=3, OR=2.12, $I^2$=58\%     & --- & --- \\
Cognition               & $k$=3, SMD=0.36, $I^2$=91\%    & --- & --- \\
Adverse events          & $k$=7, OR=1.23, $I^2$=59\%     & --- & --- \\
SSRIs depression        & --- & --- & $k$=9, SMD=$-$0.33, $I^2$=5\% \\
SSRIs cognition         & --- & --- & $k$=6, SMD=$-$0.36, $I^2$=0\% \\
SSRIs ADL               & --- & --- & $k$=5, SMD=$-$0.28, $I^2$=57\% \\
SSRIs adverse events    & --- & --- & $k$=3, OR=0.71, $I^2$=0\% \\
\midrule
\multicolumn{4}{@{}l}{\textbf{D2: CBT-I for Insomnia + Depression} (Arm C and D: 0 analyses)} \\
Depression severity     & $k$=4, SMD=$-$0.23, $I^2$=56\% & --- & --- \\
Depression remission    & $k$=4, OR=3.75, $I^2$=78\%     & --- & --- \\
Insomnia severity       & $k$=4, SMD=$-$0.32, $I^2$=76\% & --- & --- \\
Insomnia remission      & $k$=4, OR=6.47, $I^2$=53\%     & --- & --- \\
\midrule
\multicolumn{4}{@{}l}{\textbf{D3: Cholecystectomy}} \\
Mortality               & $k$=9, OR=0.12, $I^2$=29\%  & $k$=6, OR=0.14  & $k$=25, OR=0.26, $I^2$=70\% \\
Overall complications   & $k$=9, OR=0.68, $I^2$=0\%   & $k$=4, OR=0.59  & $k$=16, OR=0.52, $I^2$=34\% \\
Surgical site           & $k$=6, OR=0.18, $I^2$=0\%   & ---             & $k$=26, OR=0.26, $I^2$=18\% \\
Biliary complications   & $k$=5, OR=2.15, $I^2$=0\%   & ---             & $k$=17, OR=0.88$^*$, $I^2$=0\% \\
Intra-abdominal         & $k$=6, OR=0.49, $I^2$=0\%   & ---             & --- \\
Bleeding                & $k$=3, OR=0.57, $I^2$=0\%   & ---             & --- \\
\midrule
\multicolumn{4}{@{}l}{\textbf{D4: Pneumococcal Vaccines}} \\
PPSV23 VT-IPD           & $k$=9, VE=38\%, $I^2$=80\%  & ---                         & $k$=7, VE=40\%, $I^2$=84\% \\
PPSV23 VT-CAP           & $k$=5, VE=43\%, $I^2$=49\%  & ---                         & --- \\
PPSV23 All-cause pn.    & $k$=5, VE=38\%, $I^2$=80\%  & ---                         & --- \\
PCV13 VT-PP (obs)       & $k$=4, VE=41\%, $I^2$=3\%   & ---                         & $k$=5, VE=42\%, $I^2$=0\% \\
PCV13 VT-IPD            & $k$=2, VE=54\%, $I^2$=34\%  & ---                         & $k$=2, VE=68\%, $I^2$=0\% \\
PPSV23 VT-IPD (sub.)    & ---                          & $k$=4, VE=61\%, $I^2$=67\% & $k$=4, VE=53\%, $I^2$=81\% \\
\midrule
\multicolumn{4}{@{}l}{\textbf{D5: SGLT2i in Heart Failure}} \\
Primary composite       & $k$=6, HR=0.78, $I^2$=0\%     & ---                          & $k$=6, HR=0.78, $I^2$=0\% \\
First HHF               & $k$=7, HR=0.72, $I^2$=0\%     & $k$=7, HR=0.74, $I^2$=13\%  & $k$=7, HR=0.73, $I^2$=0\% \\
CV death                & $k$=5, HR=0.88, $I^2$=0\%     & $k$=4, HR=0.84, $I^2$=0\%   & $k$=5, HR=0.86, $I^2$=0\% \\
All-cause mortality     & $k$=6, HR=0.92, $I^2$=1\%     & ---                          & $k$=6, HR=0.91, $I^2$=0\% \\
KCCQ / QoL              & $k$=4, HR=1.98, $I^2$=99\%    & $k$=3, HR=0.93, $I^2$=99\%  & $k$=4, HR=1.98, $I^2$=99\% \\
Renal function          & $k$=4, HR=0.98, $I^2$=97\%    & ---                          & $k$=4, HR=0.98, $I^2$=97\% \\
Glycemic control        & $k$=3, HR=0.64, $I^2$=90\%    & ---                          & $k$=3, HR=0.64, $I^2$=90\% \\
\bottomrule
\end{tabular}
\end{table}

\begin{table}[ht]
\centering
\caption{Complete Phase~5 meta-analysis results (REML + Knapp--Hartung via R \texttt{metafor}). Arm~A only.}\label{tab:s4}
\small
\begin{tabular}{@{}llrllrr@{}}
\toprule
Analysis & Measure & $k$ & ES & 95\% CI & $p$ & $I^2$ \\
\midrule
\multicolumn{7}{@{}l}{\textbf{D1: Antidepressants in Elderly}} \\
All AD -- Depression     & SMD & 8 & 0.119    & [$-$0.42, 0.66]  & 0.58      & 58.5\% \\
SSRIs -- Depression      & SMD & 6 & $-$0.019 & [$-$0.73, 0.70]  & 0.95      & 50.8\% \\
Sertraline -- Response   & OR  & 3 & 2.12     & [0.20, 22.20]    & 0.22      & 57.9\% \\
Sertraline -- Cognition  & SMD & 3 & 0.364    & [$-$3.41, 4.14]  & 0.60      & 90.8\% \\
All AD -- Adverse events & OR  & 7 & 1.23     & [0.67, 2.24]     & 0.42      & 59.0\% \\
\midrule
\multicolumn{7}{@{}l}{\textbf{D2: CBT-I for Insomnia + Depression}} \\
Depression severity      & SMD & 4 & $-$0.230 & [$-$1.04, 0.58]  & 0.43      & 56.3\% \\
Depression remission     & OR  & 4 & 3.75     & [0.24, 58.84]    & 0.22      & 78.2\% \\
Insomnia severity        & SMD & 4 & $-$0.318 & [$-$1.52, 0.89]  & 0.46      & 75.9\% \\
Insomnia remission       & OR  & 4 & 6.47     & [0.86, 48.61]    & 0.06      & 52.9\% \\
\midrule
\multicolumn{7}{@{}l}{\textbf{D3: Cholecystectomy}} \\
Mortality                & OR & 9 & 0.12  & [0.05, 0.26]   & 0.0003    & 29.4\% \\
Overall complications    & OR & 9 & 0.68  & [0.55, 0.84]   & 0.0016    & 0\% \\
Surgical site            & OR & 6 & 0.18  & [0.14, 0.23]   & $<$0.0001 & 0\% \\
Intra-abdominal          & OR & 6 & 0.49  & [0.38, 0.62]   & 0.0005    & 0\% \\
Biliary complications    & OR & 5 & 2.15  & [1.38, 3.39]   & 0.009     & 0\% \\
Bleeding                 & OR & 3 & 0.57  & [0.44, 0.74]   & 0.011     & 0\% \\
\midrule
\multicolumn{7}{@{}l}{\textbf{D4: Pneumococcal Vaccines}} \\
PCV13 VT-IPD (RCT)      & OR & 2 & VE=54\% & [3\%, 96\%]  & 0.17  & 33.7\% \\
PCV13 VT-CAP (obs)      & OR & 4 & VE=41\% & [10\%, 62\%] & 0.03  & 2.8\% \\
PPSV23 VT-IPD            & OR & 9 & VE=38\% & [24\%, 49\%] & 0.0005 & 80.3\% \\
PPSV23 VT-CAP            & OR & 5 & VE=43\% & [26\%, 56\%] & 0.004 & 48.9\% \\
PPSV23 All-cause pn.    & OR & 5 & VE=38\% & [24\%, 49\%] & 0.003 & 80.1\% \\
\midrule
\multicolumn{7}{@{}l}{\textbf{D5: SGLT2i in Heart Failure}} \\
HF hospitalization       & HR & 7 & 0.724 & [0.68, 0.79] & $<$0.001  & 0\% \\
Primary composite        & HR & 6 & 0.777 & [0.73, 0.83] & $<$0.001  & 0\% \\
CV death                 & HR & 5 & 0.877 & [0.82, 0.94] & 0.003     & 0\% \\
All-cause mortality      & HR & 6 & 0.921 & [0.85, 1.00] & 0.05      & 0.7\% \\
KCCQ / QoL               & HR & 4 & 1.975 & [0.33, 11.89] & 0.28    & 99.9\% \\
Renal function           & HR & 4 & 0.980 & [0.73, 1.32] & 0.86     & 97.2\% \\
Glycemic control         & HR & 3 & 0.637 & [0.24, 1.68] & 0.17     & 89.9\% \\
\bottomrule
\end{tabular}
\end{table}

\begin{table}[ht]
\centering
\caption{Screening validation on two SYNERGY datasets at both thresholds (conservative: $\geq$4; sensitive: $\geq$3). Bold = best per dataset. One study per dataset was missed by all four arms at both thresholds.}\label{tab:s5}
\small
\begin{tabular}{@{}lllrrrrrr@{}}
\toprule
Dataset & Arm & Model & Sens $\geq$4 & Sens $\geq$3 & Spec $\geq$4 & Spec $\geq$3 & Rev.\% & Cost \\
\midrule
\textbf{Bos\_2018}   & A & Gemini & 0.500 & 0.800          & 0.892          & 0.649 & 24 & \$2.36  \\
(10 GT)               & B & GPT    & 0.600 & 0.900          & 0.893          & 0.700 & 19 & \$0.33  \\
                       & C & Claude & \textbf{0.800} & \textbf{0.900} & 0.866 & 0.691 & 17 & \$34.71 \\
                       & D & Dual   & 0.500 & \textbf{0.900} & \textbf{0.901} & 0.618 & 28 & \$2.88  \\
\midrule
\textbf{vdS\_2018}   & A & Gemini & 0.579 & \textbf{0.947} & 0.994          & 0.773          & 22          & \$17.53 \\
(38 GT)               & B & GPT    & 0.526 & 0.842          & 0.993          & 0.824          & 17          & \$2.40  \\
                       & C & Claude & \textbf{0.763} & \textbf{0.947} & 0.991 & \textbf{0.940} & \textbf{5}  & \$25.51 \\
                       & D & Dual   & 0.579 & \textbf{0.947} & \textbf{0.995} & 0.758          & 24          & \$20.02 \\
\bottomrule
\end{tabular}
\end{table}

\noindent At the conservative threshold ($\geq$4), sensitivity on van\_de\_Schoot dropped to 0.53--0.76 because the PTSD trajectory literature produced many ``undecided'' (score~=~3) judgments. The sensitive threshold ($\geq$3) absorbs these and substantially improves recall.

\end{document}